\documentclass[twocolumn]{pasj01}
\draft
\Received{2018/10/03}
\Accepted{2019/06/18}
\Published{$\langle$publication date$\rangle$}
\SetRunningHead{Y. Maeda et al.}
{X-ray telescope with an angular resolution booster}
\usepackage{color}
\usepackage{times}

\begin{document}

\title{A concept for the X-ray telescope system with an
  angular-resolution booster }

\author{
  Yoshitomo~\textsc{Maeda}\altaffilmark{1,2},
  Ryo~\textsc{Iizuka}\altaffilmark{1},
  Takayuki~\textsc{Hayashi}\altaffilmark{3,4}, 
  Toshiki~\textsc{Sato}\altaffilmark{3,4,5}, 
  Nozomi~\textsc{Nakaniwa}\altaffilmark{6},  
  Mai~\textsc{Takeo}\altaffilmark{6},  
  Hitomi~\textsc{Suzuki}\altaffilmark{6},  
  Manabu~\textsc{Ishida}\altaffilmark{1,2,6},  
  Shiro~\textsc{Ikeda}\altaffilmark{7}, 
  Mikio~\textsc{Morii}\altaffilmark{7}
}
\altaffiltext{1}{Institute of Space and Astronautical Science, Japan
  Aerospace Exploration Agency 3-1-1 Yoshinodai, Chuo-ku, Sagamihara,
  Kanagawa 229-8510}
\altaffiltext{2}{The Graduate University for Advanced Studies, 3-1-1
  Yoshinodai, Chuo-ku, Sagamihara, Kanagawa 252-5210}
\altaffiltext{3}{NASA Goddard Space Flight Center, Code 662,
  Greenbelt, MD 20771, U.S.A.}
\altaffiltext{4}{University of Maryland, Baltimore County, 1000
  Hilltop Cir, Baltimore, MD 21250 U.S.A.}
\altaffiltext{5}{RIKEN, 2-1 Hirosawa, Wako, Saitama 351-0198}
\altaffiltext{6}{Tokyo Metropolitan University, 1-1 Minami-Osawa,
  Hachioji, Tokyo 192-0397}
\altaffiltext{7}{The Institute of Statistical Mathematics, Tachikawa,
  Tokyo, 190-8562}

\email{ymaeda@astro.isas.jaxa.jp}

\maketitle

\begin{abstract} 
  We present a concept of the X-ray imaging system with high
  angular-resolution and moderate sensitivity. In this concept, a
  two-dimensional detector, i.e., imager, is put at a slightly
  out-of-focused position of the focusing mirror, rather than just at
  the mirror focus as in the standard optics, to capture the miniature
  image of objects. In addition, a set of multi-grid masks (or a
  modulation collimator) is installed in front of the telescope. We
  find that the masks work as a coded aperture camera and that they
  boost the angular resolution of the focusing optics. The major
  advantage of this concept is that a much better angular resolution
  an order of 2--3 or more than in the conventional optics is
  achievable, while a high throughput (large effective area) is
  maintained, which is crucial in photon-limited high-energy
  astronomy, because any type of mirrors, including light-weight
  reflective mirrors, can be employed in our concept. If the
  signal-to-noise ratio is sufficiently high, we estimate that angular
  resolutions at the diffraction limit of 4$''$ and 0\farcs4 at
  $\sim$7~keV can be achieved with a pair of masks at separations of
  1~m and 100~m, respectively, at the diffraction limit.
\end{abstract}

\section{Introduction}

Focusing incoming X-rays with reflective X-ray telescopes is by far
the best way of imaging of astronomical objects, as well as achieving
a much higher sensitivity than is otherwise possible.  Historically,
the total reflection technology was used at first (see
\cite{2009ExA....26...95A} for a review). The Bragg reflection or
diffraction was then used , being considered as an effective method to
extend the energy band beyond 10 keV, called the supermirror
technology in the hard X-ray band (e.g., \cite{1995ApOpt..34.7935J},
\cite{1998ApOpt..37.8067Y}) and a Laue lens in the soft $\gamma$-ray
band (e.g., \cite{1982RScI...53..131S}).  Since the focusing optics
makes incoming photons concentrated, the focal plane detectors can be
much smaller than is otherwise possible.  Focusing makes a
signal-to-noise ratio increase significantly, or even dramatically
because the particle background in space is proportional to the area
or volume of the detecting area in the detector. The high
signal-to-noise ratio makes the sensitivity improve by many orders,
compared with non-focusing detectors.

A small detector can also have good energy resolution (e.g.,
\cite{1991ITED...38.1069B} for X-ray CCDs; \cite{2007PASJ...59S..77K}
for calorimeter arrays) and polarimetric sensitivity (see
\cite{2016SPIE.9905E..17W}). The focusing optics is therefore a key
technique to realize imaging with high-resolution spectroscopy and
potentially even with polarimetry.

\begin{figure}[h]
  \begin{center}
    \includegraphics[width=0.85\textwidth]{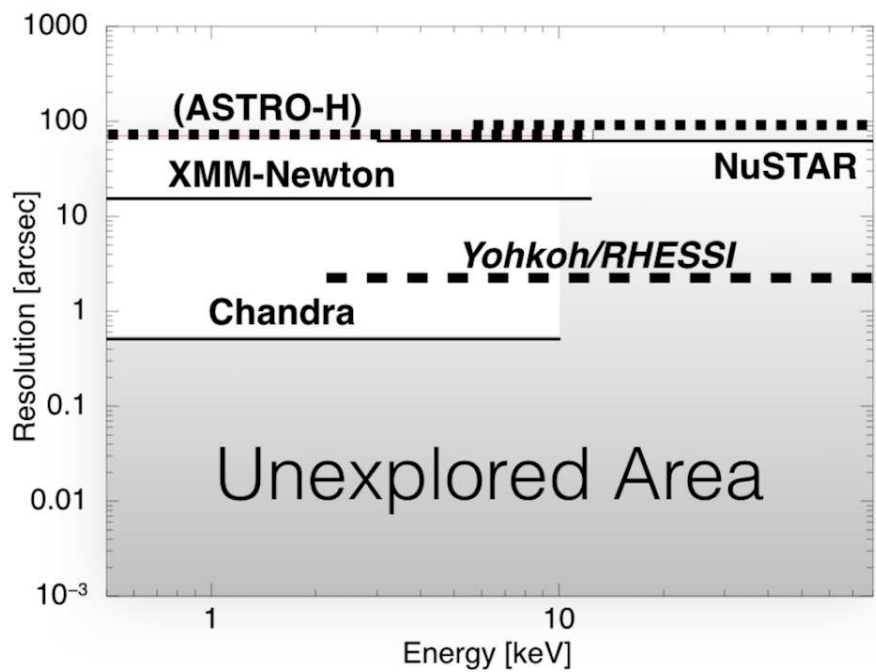}
    \vspace*{-0.3cm}
  \end{center}
  \caption{Angular resolution in half-power diameter of several X-ray
    satellites. The best angular resolution (solid line) of the
    focusing optics is archived by Chandra at $<$10~keV, XMM-Newton at
    $\sim$10--16 keV \citep{2000SPIE.4012..731A}, and NuSTAR at
    $\sim$16--78 keV \citep{2015ApJS..220....8M}. For comparison, that
    of ASTRO-H, which employs typical lightweight mirror modules is
    also plotted by a dotted line. The dashed line shows the
    resolution of the non-focusing optics (Yohkoh and RHESSI).}
  \label{fig:angularresolution}
\end{figure}

Since the X-ray flux from celestial objects is usually low, to achieve
the largest possible effective area in the given restriction is vital
for developing focusing optics (see \cite{2009ExA....26...95A} for a
review).  Due to the tight limit of the total weight of a spacecraft,
light-weight mirror modules have been the primary focus of development
in the field worldwide.  Indeed, light-weight mirror modules with a
large effective area have been deployed in X-ray satellites (e.g.,
\cite{1995PASJ...47..105S}).  However, none of those light weight
mirror modules has achieved a better angular resolution than that of
the Chandra High Resolution Mirror Assembly (HRMA), $\sim 0.5$ arcsec,
which consists of a set of mirror assemblies that are heavy and have a
comparatively low effective area, but they maintain ultimate
smoothness and precision in the surface shape of the mirrors.  It
would be very difficult to achieve a better angular resolution than
that of the Chandra HRMA without sacrificing the effective area, and
vice versa, unless an innovation or breakthrough could somehow be made
in the X-ray telescope technology (e.g.,
\cite{2018SPIE10699E..0OZ}). Accordingly, observations with a high
angular-resolution at arcseconds or better to match those in other
wavelengths, such as the optical band, remain an unexplored area in
high-energy astronomy (figure \ref{fig:angularresolution}).

For example, a molecular cloud near the center of a galaxy is
populated in the unexplored area. By using the angular resolution of
$\sim$3 arcmins, \citet{1996PASJ...48..249K} discovered the
fluorescent X-ray emission from cold iron atoms in molecular clouds
from the center of our Galaxy. The line is possibly due to irradiation
by X-rays from a hidden central engine, which was bright in the past,
but is presently dim. To detect similar emission from an external
galaxy, sub-arcsec resolution would be needed: for example 0.1 arcsec
for NGC 1068 at the distance of 14.4 Mpc. High angular resolution
observations in an unexplored area may open a new window for studying
structure near the center of galaxies.

Chandra's arcsecond resolution gives us a spatially resolved image of
the shell of supernova remnants, such as Cas A
(\cite{2000ApJ...528L.109H}). The time variation of the X-ray images
is a probe of particle acceleration at the shell
(\cite{2008ApJ...677L.105U,2018ApJ...853...46S}). The hard X-ray band
above the Chandra energy band (i.e., $> 10$ keV) is likely dominated
by non-thermal emission
(\cite{2009PASJ...61.1217M,2013ApJ...770..103H}). The arcsecond
resolution image at 10 keV or above would also be a new tool to
address particle acceleration at the shell.

\citet{1987ApOpt..26.2915C} presented the idea of out-of-plane or
off-plane imaging that can lead to the high-resolution observations
without exceeding the current state of the art in optical
fabrication. He pointed out that both scattering and figure errors in
grazing incidence optics are larger in the plane of incidence than out
of plane by a factor equal to $1/\sin\Theta$, where $\Theta$ is the
grazing angle.  When the full annular aperture of a grazing incidence
mirror module is stopped down, the point spread function becomes
highly elliptical with a width as much as $\sin \Theta$ times narrower
than the full image, indicating that improvements in resolution of up
to 100 times can be achieved.  Such an improvement leads us a high
resolution in one dimension.  By observing the source at a series of
orientations, full two-dimensional image information can be retrieved.

Several X-ray astronomy missions using this out-of-plane imaging technique have been proposed: the latest is Arcus (\cite{2016SPIE.9905E..4MS}). Arcus is not an imaging mission but an x-ray grating spectrometer explorer. In this mission, 
an improvement of the angular resolution grazing incidence in the out-of-plane direction is used to focus a narrow width in the dispersion direction of the grating spectrometer. A high angular resolution of $\sim$1.4 arcsecs in half power width (HPW) along the out-of-plane direction was measured for the mirror unit (\cite{2017SPIE10399E..0CC}). The angular resolution also provides high-dispersion spectra with superior energy resolution. Again, in order to achieve the arcsecond resolution along the out-of-plane direction, the mirror had a moderately good resolution of $\sim14$ arcsec in half-power diameter (HPD). 

 As a completely different method, X-ray optics using the diffraction or interferometry has been discussed. 
For example, \citet{2008SPIE.7011E..0TS} proposed the MASSIM mission, in which diffractive and refractive focusing optics with a 1~m diameter is employed. 
By combining six Fresnel lenses, it covers an energy band of 4.5--11 keV. In their proposed design, MASSIM achieves a milli-arcsecond resolution with a large effective area of a few thousand cm$^2$.
\citet{2005AdSpR..35..122C} proposed an interferometer that uses flat collecting mirrors, whereby a milli-arcsecond angular resolution or better can be realized. 
For example, one of the proposed designs, called the MAXIM Pathfinder, covers the 0.5--2 keV plus 6 keV.

At present, the only existing
optical system that may surpass reflecting focusing optics in angular
resolution is shadow mask technology. 
This employs a flat aperture-patterned grille in front of the detector. 
The pattern works as a multiple pin-hole camera. 
The detected pattern is then reconstructed by post-processing of the signal.
A spatial resolution of as high as a few arcsecs over a wide energy range
has been achieved (\cite{1991SoPh..136...17K}, \cite{2002SoPh..210....3L}).
In principle, the mask technology can achieve much better angular resolution if we make a longer separation of the masks
(\cite{1988SoPh..118..269P}, see also eq. \ref{eq4} ).
For example, the angular resolution at 7 keV is 0.04 and 0.004 arcsecs for the separations of 100~m and 10~km, respectively. 
However, since it does not focus the incoming light, its sensitivity
is orders of magnitude worse than that of the focusing optics
due to high background.
Its practical application is thus limited to the brightest objects such as the Sun.

In this paper, we present a new concept to combine focusing optics and mask technologies. In this concept, high sensitivity and photon collimation by the focusing optics and 
 high angular-resolution by the coded-mask are simultaneously attained. This concept gives a wide variety of mission design for various energy ranges with various combinations of focal-plane detectors. It is notable that the angular resolution of the mirror assembly of the focusing optics is not required to be better than that of the masks and is acceptable even on an order of arcminutes. 

\begin{figure}[ht]
 \begin{center}
  \FigureFile(85mm,85mm){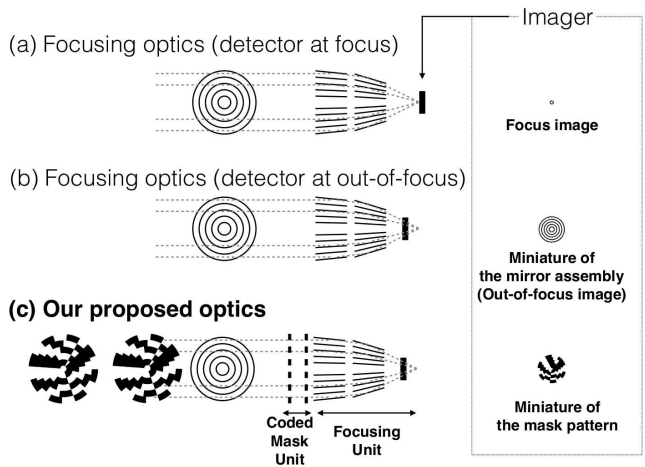}
 \end{center}
 \caption{
Concept of the focusing optics system with an angular resolution booster (c). The mask units are mounted in front. 
For comparison, the normal telescope system is displayed. A two-dimensional detector is placed at (a) the focus  and (b) an out-of-focus position. }
\label{fig:gaiyo}
\end{figure}
\begin{figure}[p]
 \begin{center}
  \FigureFile(75mm,75mm){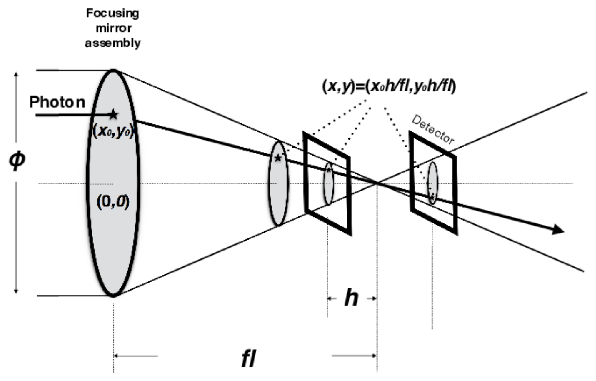}\\
    \vspace*{-.4cm}
 \FigureFile(55mm,55mm){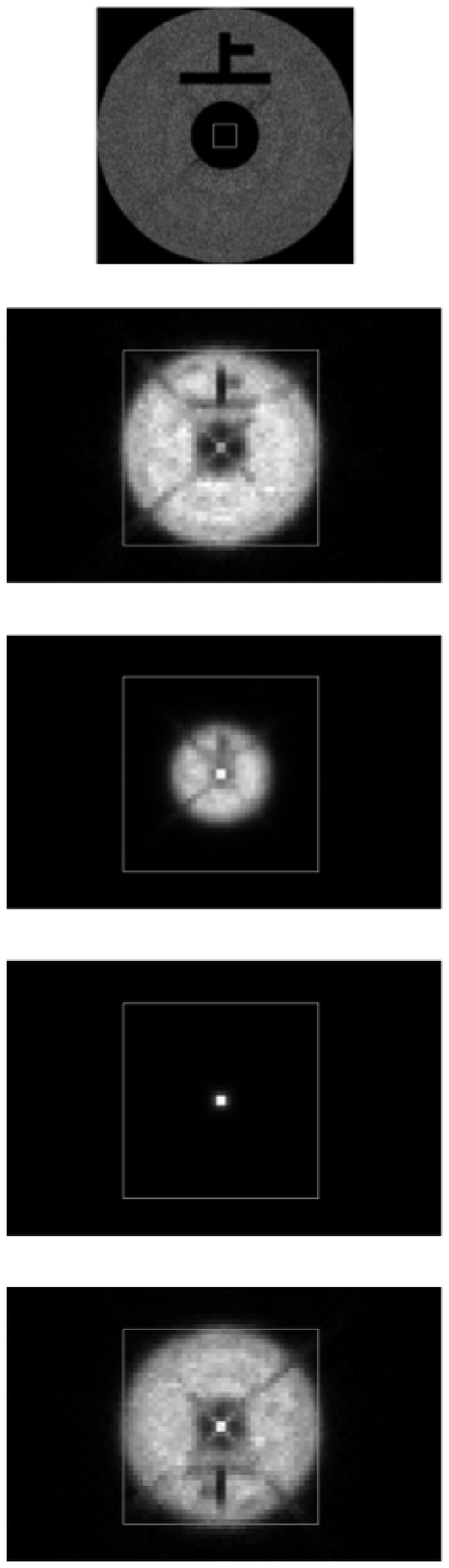}\\
  \end{center}
 \caption{
Example of a miniature effect of focusing optics with a mask of the Chinese/Japanese character ``\includegraphics[width=1.0em]{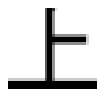}''. The top panel shows a schematic view, and the others, from the second to bottom panels, shows the photon maps at the mouth of the mirror assembly ($h=+5600$ mm), and at $h=+500$ mm, $+250$ mm, 0 mm and $-500$ mm from the focal length. The 40mm square is overlaid for reference. }
\label{fig:miniature}
\end{figure}

\section{Telescope concept}

Figure~\ref{fig:gaiyo}c shows the concept of the proposed telescope system. The system consists of a mask unit in front and a focusing unit behind, the latter
used for collimating photons. 
In the configuration of the standard optics, a two-dimensional detector (i.e., imager) is placed at the best focus, i.e., at the focal length (figure~\ref{fig:gaiyo}a). 
 In our concept, in contrast, it is put at a slightly out-of-focus position (figure~\ref{fig:gaiyo}b). 
In addition, one or a set of multiple grid masks (or modulation collimators) is installed in front of the telescope (figure~\ref{fig:gaiyo}c). 
The multi-grid masks work as a coded aperture camera and behave like an angular-resolution booster of the focusing optics. 

In figure~\ref{fig:gaiyo}-c, we show the double slits -- the primary and secondary mask -- for the mask unit. The secondary can be removed, as with usual coded-masks. 
For a single slit, the boundary of each pixel of its imager or each detector of its array is known to work as a secondary slit. The structure of the focusing mirror assembly can also work as a secondary slit. For simplicity, we assume a double slit here and elsewhere. 

In figure~\ref{fig:miniature}, we present an example of a mask pattern ``\includegraphics[width=1.0em]{figs/ue.eps}'', which is a representive of Chinese/Japanese character meaning ``upper''. 
We assume the same pattern for the primary and secondary masks. The incoming photon moves along the on-axis direction . 
The mask pattern shrinks toward the focus, and becomes a dot-like at the focus. It stretches again behind the focus, but with a double flip (i.e., $180^\circ$ rotation). 
In other words, the mask pattern is preserved in miniature at the image on out-of-focus planes, but is blurred with the angular resolution of the focusing mirrors. 
This means that the imaging technique with the multi-grid mask can be applied by using the miniature image at an out-of-focus position of the focusing optics. 
It is notable that the image at the focus has almost no information on the mask pattern. Therefore, the detector must be located at the out-of-focus position\footnote{Exceptionally, the detectors can be located at the focus if multiple focusing units (i.e., a mirror module array) are considered and if the different mask pattern is assigned for each focusing unit. In this exceptional case, each telescope works as each detector array of the masks. }. 
The advantage of using a miniature image is discussed in section~3. 

The mathematical formulation of this concept is described as follows.
The count-rate distribution and the X-ray flux spatial distribution are
two-dimensional images on an imager and on the sky, respectively.
We rearrange them into vectors, {\bf R} and {\bf F}.

In the standard mask technology, the count-rate distribution ${\bf R}$ is given by
 \begin{eqnarray}                          
{\bf R} = M {\bf F} + {\bf N},
  \label{eq0a}                            
 \vspace{-3cm}
  \end{eqnarray}
where ${\bf M}$ is the matrix of the encoding pattern of the masks,
and ${\bf N}$ is a vector of the non-X-ray background spatial distribution.
The X-ray flux spatial distribution, F, is estimated by a reconstruction technique.


 In our concept, we introduce the response matrix $\it{T}$ of the focusing mirror assembly between the masks and detector (figure~\ref{fig:gaiyo}).
The equation \ref{eq0a} is then modified as      
 \begin{eqnarray}                          
{\bf R} = T M {\bf F}+ {\bf N}. \end{eqnarray} 
\label{eq0d}                          
For the ideal angular resolution (i.e., 0 arcsec), both $\it{T}$ is just a constant linear matrix to describe the miniature image. 
In reality, the matrix  $\it{T}$ is complicated and cannot be modeled analytically. 
The matrix must be constructed by detailed measurements (e.g., \cite{2008JaJAP..47.5743O}; \cite{2016JATIS...2d4001S}). 
The spatial distribution {\bf F} can then be estimated by reconstruction (see our companion paper \cite{2019PASJ...71...24M}).

\section{Discussion}

\subsection{High angular-resolution bonus with the mask unit}

As demonstrated in our companion paper \citet{2019PASJ...71...24M}, the angular resolution is given by the angle subtended by the first mask element, as seen from the second element ($\rm \bf M$ in Eq. \ref{eq0d}).  When the mask elements are squared with size of 
$s$, and when the separation of the two grid masks is $d$, 
the angular resolution $\theta_{\rm mask}$ (FWHM) along each axis at an off-axis angle, $\tau$, in the observed sky is equal to:

\begin{eqnarray}                          
  \Delta \theta_{\rm mask} = {\rm arctan}(\frac{s\ {\rm cos}^2 \tau}{d}). 
  \label{eq1}                            
 \end{eqnarray} 
 The factor of the square of cosine is due to projection effects.  In the case of a narrow field of view ($\tau\ll1$), it is approximated as 
  \begin{eqnarray}                          
  \Delta \theta_{\rm mask} \approx s / d. 
  \label{eq2}                            
 \end{eqnarray}                             
 
At the diffraction limiting, 
  \begin{eqnarray}                          
     \Delta \theta_{\rm mask} = 2\lambda/s, 
  \label{eq3}                            
 \end{eqnarray}      
where $\lambda$ is the photon wavelength. 
Using equations \ref{eq2} and \ref{eq3}, the well-known formula of  
  \begin{eqnarray}                          
\Delta \theta_{\rm mask} = \sqrt{\frac{2\lambda}{d}}
  \label{eq4}                            
 \end{eqnarray}     
is obtained in the case that the angular resolution is limited by diffraction (\cite{1965ApOpt...4..143O}). 

\begin{figure}[ht!]
 \begin{center}
   \includegraphics[width=0.45\textwidth]{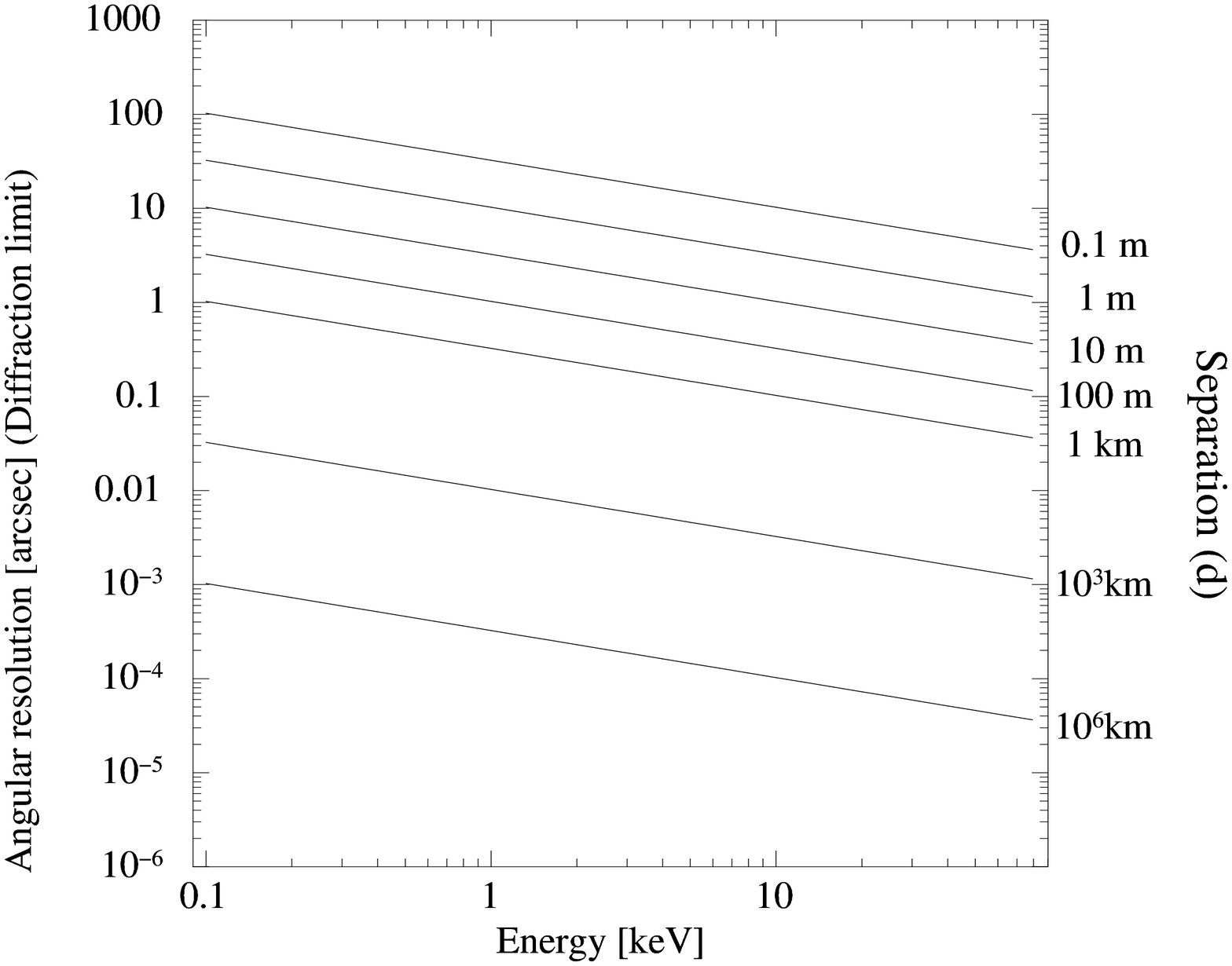}
 \end{center}
 \caption{Diffraction limit for the angular resolution of a mask.
}
\label{fig:diffractionlimit}
\end{figure}

Figure \ref{fig:diffractionlimit}
shows the diffraction limit angle $\Delta \theta_{\rm mask}$ versus energy, $E$, for several different grid separations for a pair of slits (e.g., \cite{1988SoPh..118..269P}). 
At 7 keV for the iron-K emission lines, $\Delta \theta_{\rm mask}$ $=$ 4 and 0.4 arcsecs can be achieved for two distances of 1~m and 100~m, respectively. 
These angular resolutions are orders of magnitude better than those obtained with the light-weight focusing optics ($\Delta \theta_{\rm mirror} $ $\approx$ 1 arcmin in HPD), such as ASTRO-H (SXT: \cite{2016SPIE.9905E..0ZO}, HXT: \cite{2014ApOpt..53.7664A}). 


 The angular resolution is also affected by the other factors such as the angular resolution of the telescope (i.e., the matrix $T$) and the signal-to-noise ratio of the images (the errors of the vectors $\bf \rm R$ and $\bf \rm N$). 
 It is because the count-rate distribution $\bf \rm R$ is blurred by the matrix $\bf \rm T$ (e.g., figure \ref{fig:miniature}, the encoding pattern is blurred). 
The distribution $\bf \rm R$ fluctuates because of photon statistics or noise $\bf \rm N$. 
Quantitative estimation of these pattern recognition errors will be future work. 
An influence of the recognition error due to the photon statistics is briefly commented on \citet{2019PASJ...71...24M}.

\subsection{Sensitivity benefit from the focusing unit}

 Since the focusing mirror assembly works to produce a miniature image of incoming photons, the sensitivity at the detector is greatly improved from what is expected from a simple mask. 
In this section, we discuss the reduction factor, $f_{\rm bgd}$, of the detector background of our telescope system to that of the mask. We assume that the geometrical area of the mirror assembly is the same as that of the mask. 

Let us introduce an out-of-focus factor, $\alpha$, as
\begin{eqnarray}  
 \alpha =  \frac{ {\frac{h}{fl}} \ \phi }{ fl \ \Delta \theta_{\rm mirror} }
\end{eqnarray}      
where $h$, $fl$, and $\phi$ are 
the out-of-focus distance from the focus, the focal length, and the diameter of the focusing mirror assembly (see figure~\ref{fig:miniature}). The denominator and numerator represent the image size at the focus position (or the angular resolution of the focusing optics) and that at the out-of-focus, respectively. 
The angular resolution, $\Delta \theta_{\rm mirror}$, should be good enough to resolve the mask pattern that appears to be the miniature image at an out-of-focus position (i.e., $\alpha \gg 1$). 

The reduction factor of the detector background can be given by 
  \begin{eqnarray}                         
     f_{\rm bgd} = 
      (\frac{A_{\rm geometry}}{A_{\rm mirror}}) (\frac{h}{fl})^2 
       \label{eqsen}                            
 \end{eqnarray}      
 or
  \begin{eqnarray}                         
     f_{\rm bgd} = 
      (\frac{A_{\rm geometry}}{A_{\rm mirror}}) ( \frac{\alpha \ \Delta \theta_{\rm mirror} \ fl}{\phi})^2 ,
       \label{eqsen13}                            
 \end{eqnarray}      
where $A_{\rm mirror}$ is the effective area of the focusing mirror assembly, and $A_{\rm geometry}$ is the geometrical area of the mirror assembly or the mask.
In the standard design of the focusing optics, the tilt of the outer diameter is set to be the critical angle $\theta_{\rm cri}$. 
For double reflection, such as in the Wolter I optics, 
   \begin{eqnarray}  
    \theta_{\rm cri} = \phi / 8 fl,
       \label{eqsenwolter}                            
 \end{eqnarray}      
 and therefore,
   \begin{eqnarray}                         
     f_{\rm bgd} = 
      (\frac{A_{\rm geometry}}{A_{\rm mirror}}) ( \frac{\alpha \ \Delta \theta_{\rm mirror}}{8\ \theta_{\rm cri}})^2 .
       \label{eqsen23}                            
 \end{eqnarray}

\begin{figure}[ht!]
 \begin{center}
    \includegraphics[width=0.45\textwidth]{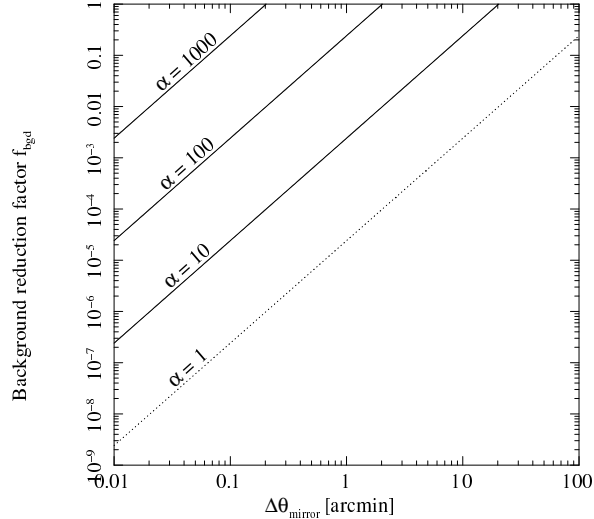}
 \end{center}
 \caption{Limit of the reduction factor $f_{\rm bgd}$ of the detector background.  A Wolter-I focusing optics optimized for 7 keV is assumed (see text for details). 
 The cases of the out-of-focus factor $\alpha$ = 1, 10, 100, and 1000 are plotted.  Note that the case of $\alpha$ = 1 is for the detector at the focus and for reference only. }
\label{fig:backgroundreductionfactor}
\end{figure}

Figure~\ref{fig:backgroundreductionfactor} shows the limits of the background reduction factor, $f_{\rm bgd}$, for $\alpha$ = 10, 100, and 1000. 
We assumed that the throughput ratio ${A_{\rm mirror}}/{A_{\rm geometry}}$ of the focusing optics is 0.5 and the critical angle $\theta_{\rm cri} = 0.6$~degree, which corresponds to the critical angle of a gold surface at 7 keV. The $f$-number of the Wolter-I optics with 0.6 degrees tilt at the outer diameter is $\sim$0.08, which is similar to those of the XMM-Newton, Suzaku, and Hitomi soft X-ray mirror modules. 
 
A better reduction factor, $f_{\rm bgd}$, is obtained for a smaller out-of-focus factor, $\alpha$, and a better angular resolution of the focusing optics.
The out-of-focus factor, $\alpha$, should be  optimized well in the design. 
We find that  $f_{\rm bgd}=10^{-3}\sim10^{-1}$ is obtained for $\alpha=10$--$100$ for an angular resolution of the focusing optics of 1 arcmin (e.g., ASTRO-H).
Even in a much worse case of 10-arcmin resolution for the focusing optics,
$f_{\rm bgd}=10^{-1}$ is obtained for $\alpha=10$. 

\subsection{A simulated example}

We now demonstrate the merits discussed above based on a Monte-Carlo simulation combined with a reconstruction technique. For this, what we need to assume is the matrix of the encoding pattern of the masks $M$, and the response matrix $\it{T}$ of the focusing-mirror assembly. A zero vector for the background spatial distribution, ${\bf N}$, is assumed. The technique for the reconstruction can be referred in our companion paper \citet{2019PASJ...71...24M}.

\begin{figure}[b]
\begin{center}
\includegraphics[width=0.67\textwidth]{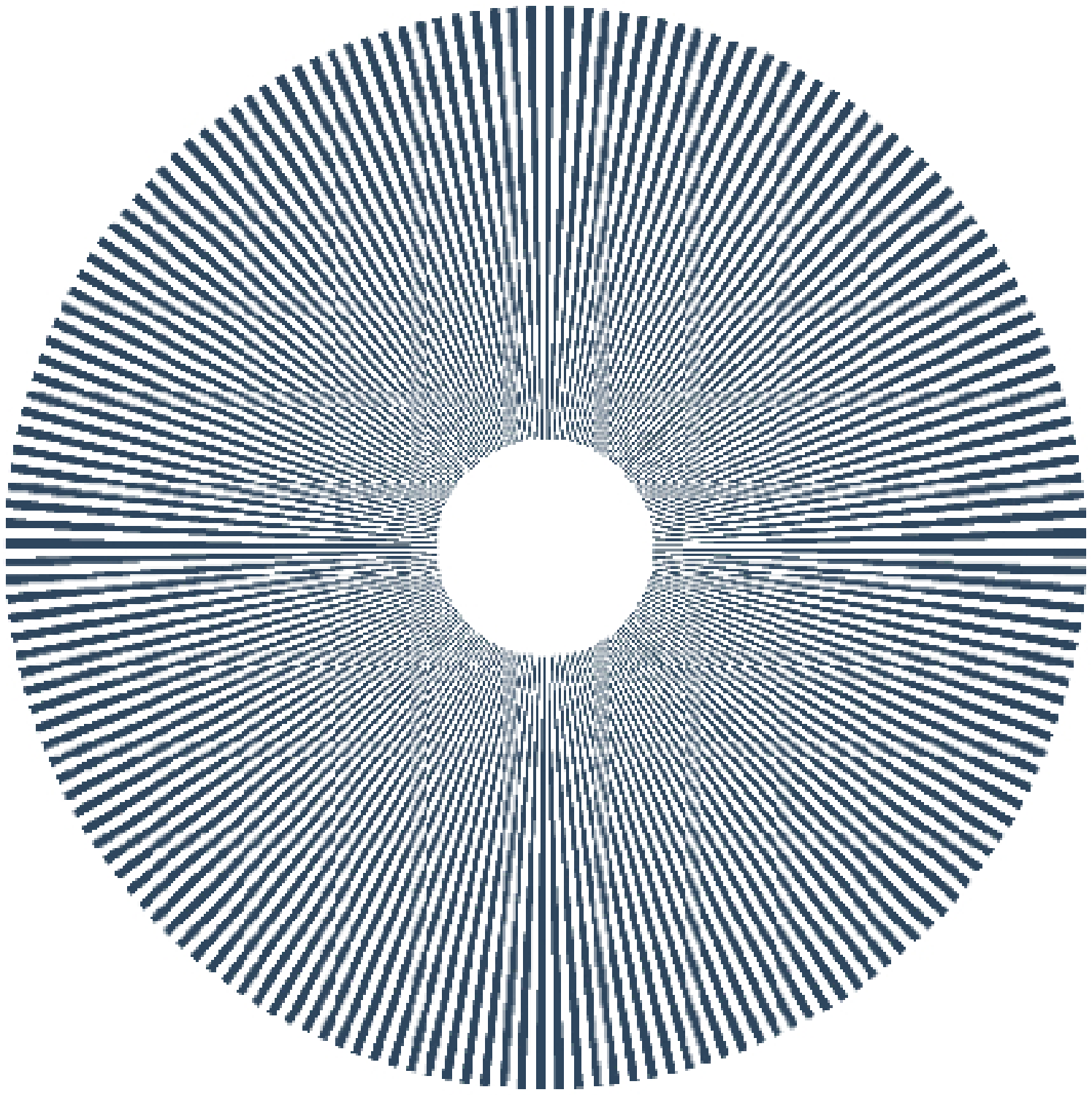}
 \end{center}
 \caption{Schematics of a rotation mask for the simulation. One radial slit appears at every 0.025 degrees.
 The opening fraction is 50 percent.
The secondary slit is designed to mask half of photons toward the on-axis direction. 
The slit sizes, $s$, of the innermost and outermost radii are 25 $\mu$m and 98$\mu$m, respectively. 
In total, 7200 slits are opened in each mask.
 Two masks are placed at a distance 
of 1~m. 
 The second mask is rotated by one quater of the pattern. 
}
\label{fig:slitexample}
\end{figure}

The matrix $M$ is calculated for the radial double mask given in figure \ref{fig:slitexample}. 
One radial slit appears at every 0.025 degrees.
 The opening fraction is 50 percent.
The secondary slit is the same but is aligned so as to mask half of the photons toward the on-axis direction. 
The slit sizes, $s$, of the innermost and outermost radii are 25 $\mu$m and 98 $\mu$m, respectively. 
In total, 7200 slits are opened in each mask.
 Two masks are placed with a distance, $d$, of 1~m. 
 The second mask is rotated by one quater of the pattern. 
 
The 25 $\mu$m and 98$\mu$m size correspond to the 5 and 20~arcsecs angler scale for the 1~m distance; 
5~arcsecs is just at around the diffraction limit at 7 keV ($\sim$4 arcsec: figure \ref{fig:diffractionlimit}); 
20~arcsecs is about one-third of the HPD of the angular resolution of the focusing mirror assembly that we assumed (see below). 
The slit design is sensitive to the angular scale from 1/10 to 1/3 of it. 
The mask is assumed to be zero of thickness and opaque for X-rays.

The matrix $\it{T}$ of the focusing mirror assembly is calculated with a ray-tracing program “xrtraytrace” (Yaqoob et al. 2019 in prep.). The optics parameter of an ASTRO-H SXT-like light weight mirror module (\cite{2016SPIE.9905E..0ZO}) is adopted in the simulation. 
The diameter of the module is 45~cm whereas the focal length is 5.6~m. 
The angular resolution we set for this simulation is about 1 arcmin. 

By using the Monte Carlo method, the position of the first slit of the incident photon is randomly given. For each photon, we calculate the raypath to the out-of-focus position. This calculation was made by input offset angles with a 2 arcsec pitch. The arrival photon map was made for each offset. 
We then obtained a matrix of the mask and the focusing units ($T M$) with a 2~arcsec pitch. 

Examples of the out-of-focus images are given in figure \ref{fig:slitexample}. Each image was extracted at the height $h=$500mm from the focal length. Since the focal length of the SXT is 5.6~m, the image roughly corresponds to the miniature image of the mirror module structure by an order of $\alpha\sim10^{-1}$. 
An image at each off-axis angle gives a unique pattern. These patterns can be used to reconstruct the images.  
Since we binned the images by 1mm, the size of the matrix $T M$ is $3600 \times 3600$. 
With a similar way, we ran the simulation by inputting celestial objects. 
Detailed reconstruction procedures can be found in our companion paper by \citet{2019PASJ...71...24M}. 

\begin{figure}[ht!]
 \begin{center}
\includegraphics[angle=0,width=0.45\textwidth]{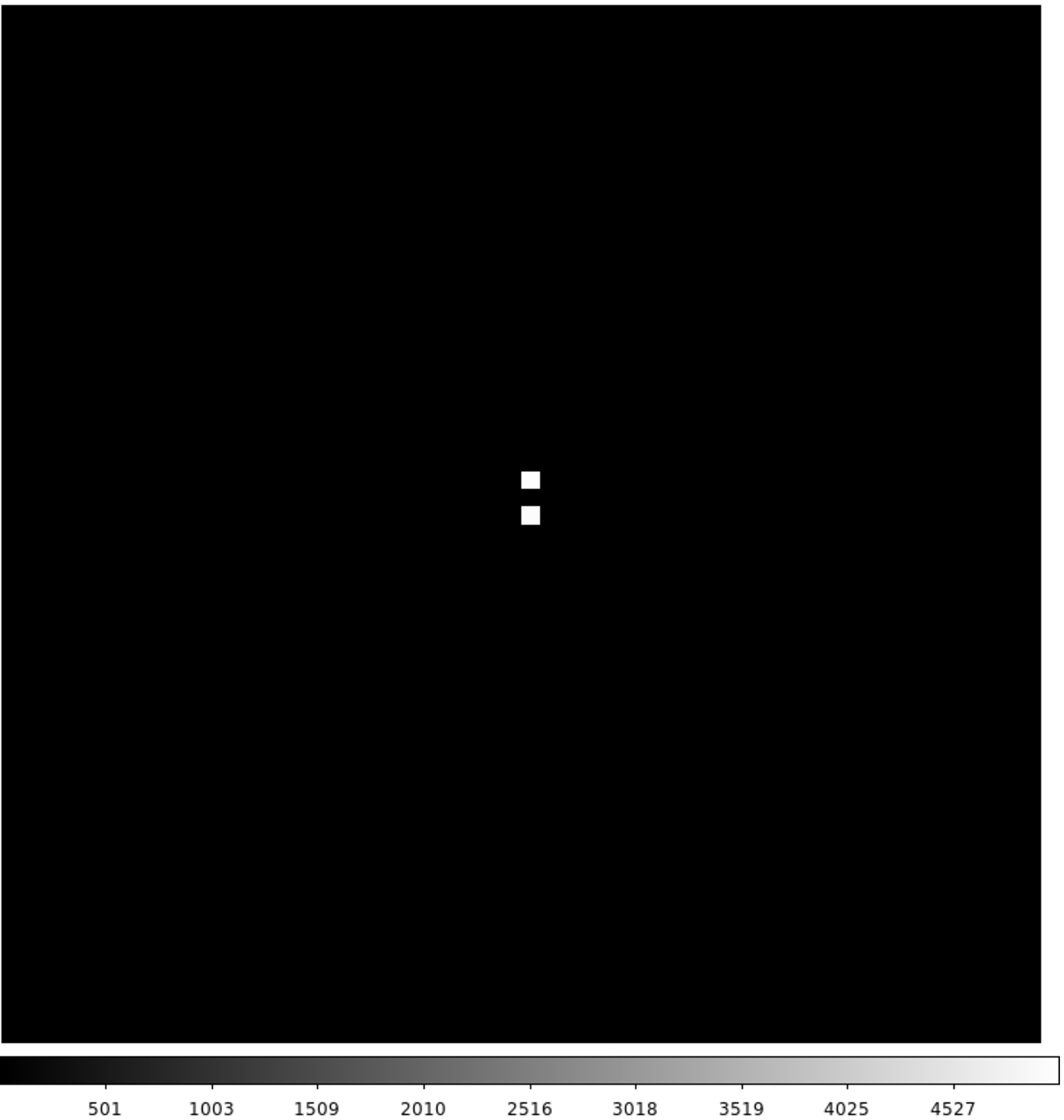}
\includegraphics[angle=0,width=0.45\textwidth]{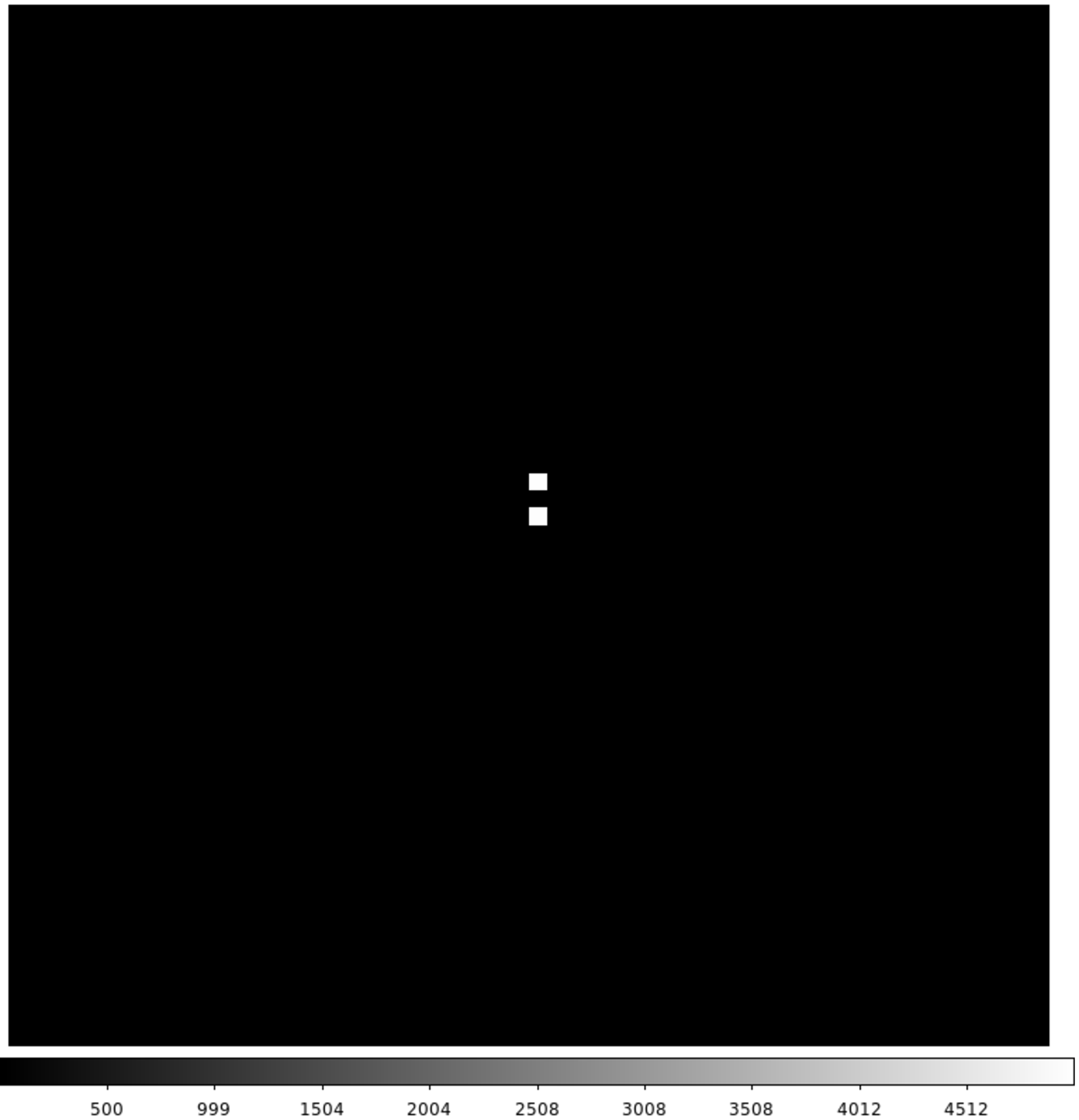}
 \end{center}
 \caption{Image reconstruction for a double star. The left and right
   panels correspond to the input and the reconstructed image for a
   double star with a distance of 4 arcsecs. They are reprinted of the
   panel (a) and (d) of Figure 1 in \cite{2019PASJ...71...24M}.}
\label{fig:doublestar}
\end{figure}

\citet{2019PASJ...71...24M} clearly resolve the double point sources separated by 4 arcsecs with $5\times10^3$ photons (figure \ref{fig:doublestar}). 
Celestial objects similar to the double or multiple point-like sources can be found in an imaging stellar binary 
(e.g., TWA 5: \cite{2003ApJ...587L..51T}, WR 147: \cite{2010ApJ...721..518Z}), an binary black hole (e.g., NGC 6240: \cite{2003ApJ...582L..15K}) and a gravitational lensed quasar (e.g., RX~J$1131-1231$: \cite{2012ApJ...757..137C}). Objects with these simple structures could be promising targets for the Angular Resolution Booster (ARB). 

\begin{figure}[ht!]
  \begin{center}
    (a) Image at focus \hspace{1.2cm}
    (b) Image at out-of-focus \hspace{0.6cm} \\  \vspace*{-0.6cm} 
    \FigureFile(40.0mm,40.0mm){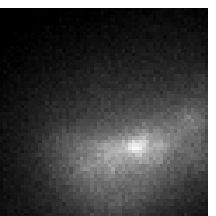}
    \hspace{0.6cm} \vspace{-0.6cm}
    \FigureFile(42.0mm,53.0mm){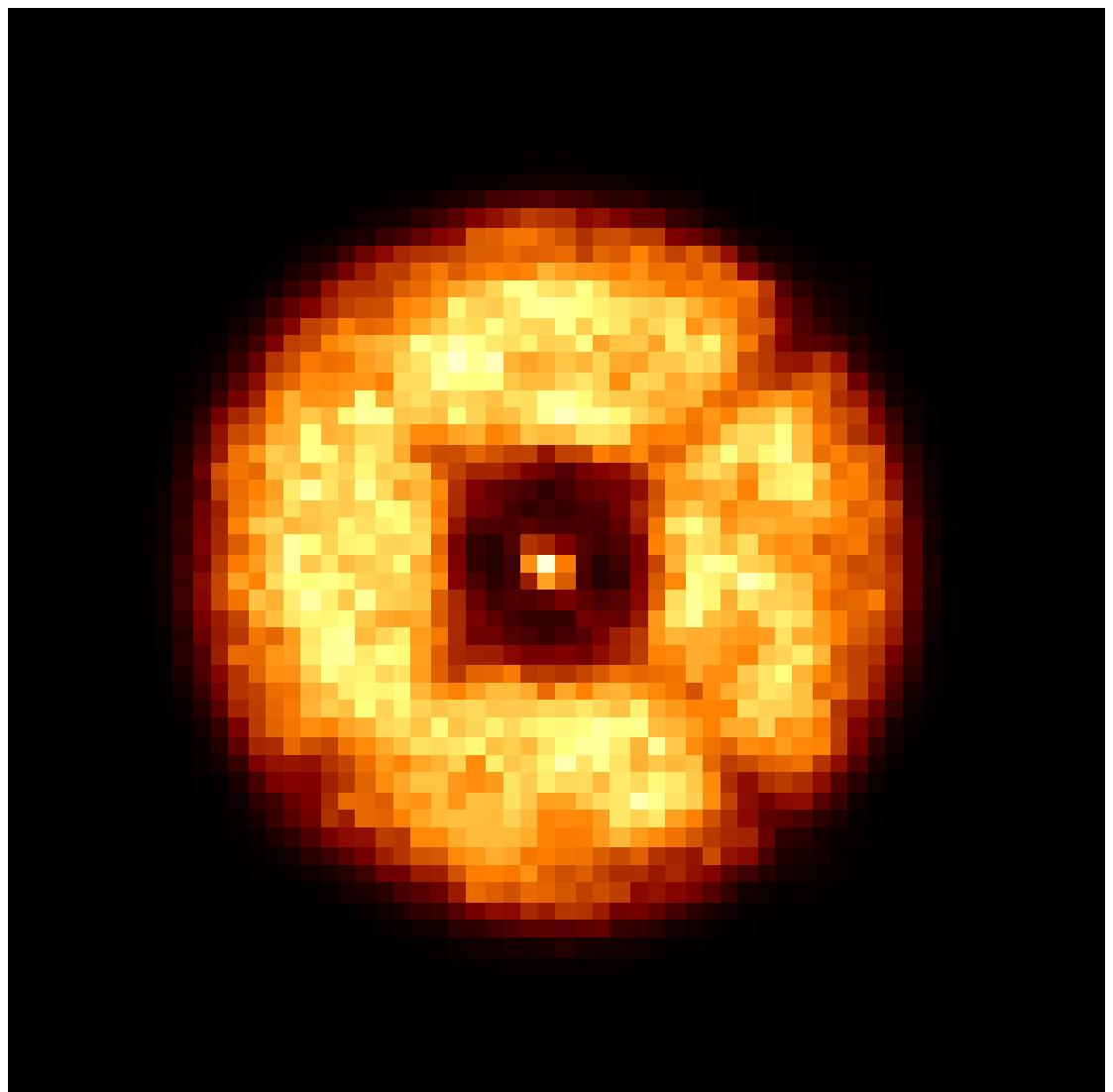}\\
    \vspace*{1.0cm}
    (c) Reconstructed images using the out-of-focus images\\
    1M,   \hspace*{2.8cm}  10M,   \hspace*{2.8cm} 100M\\
    \vspace*{-0.70cm}
    \FigureFile(40.0mm,40.0mm){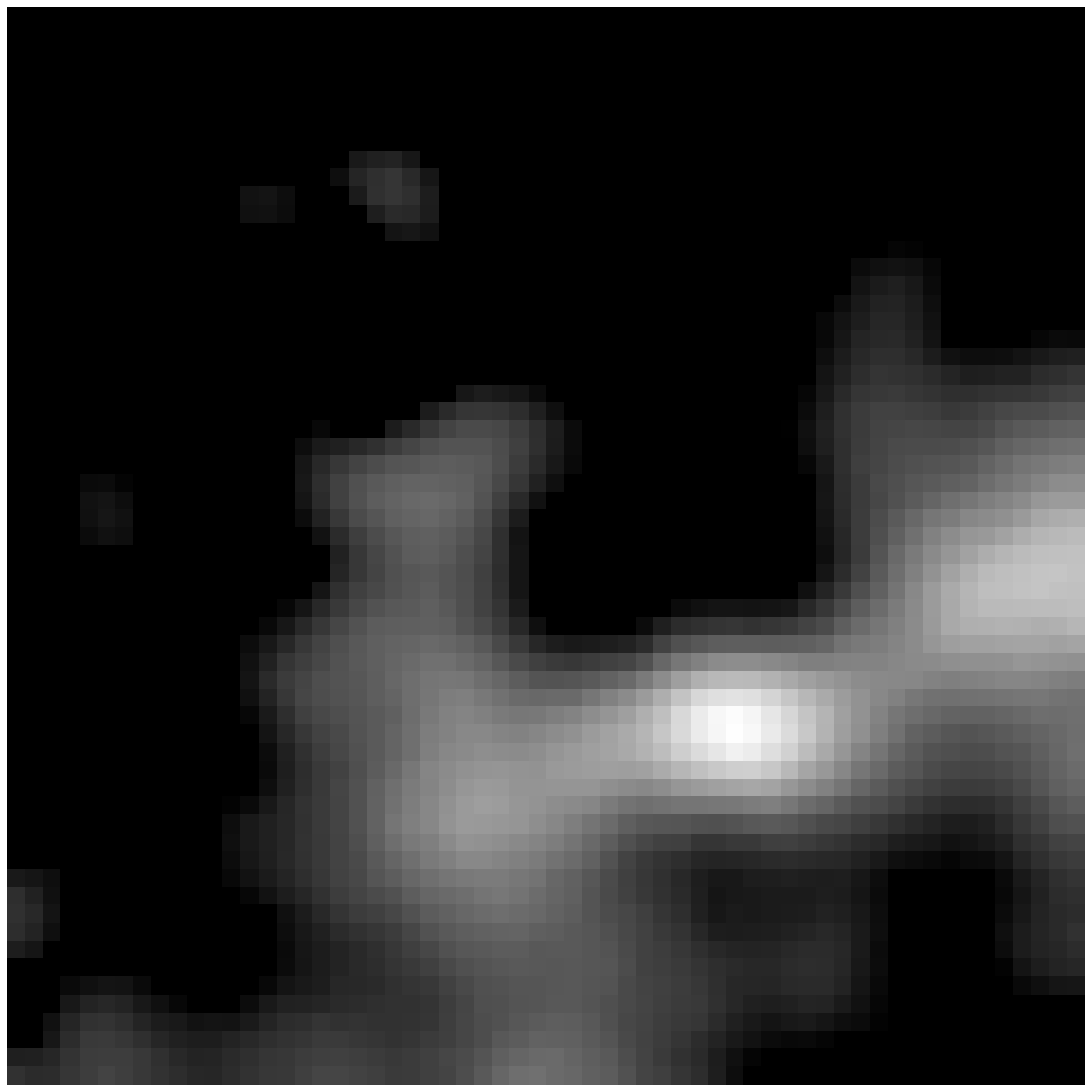}
    \hspace*{1.0cm}
    \FigureFile(40.0mm,40.0mm){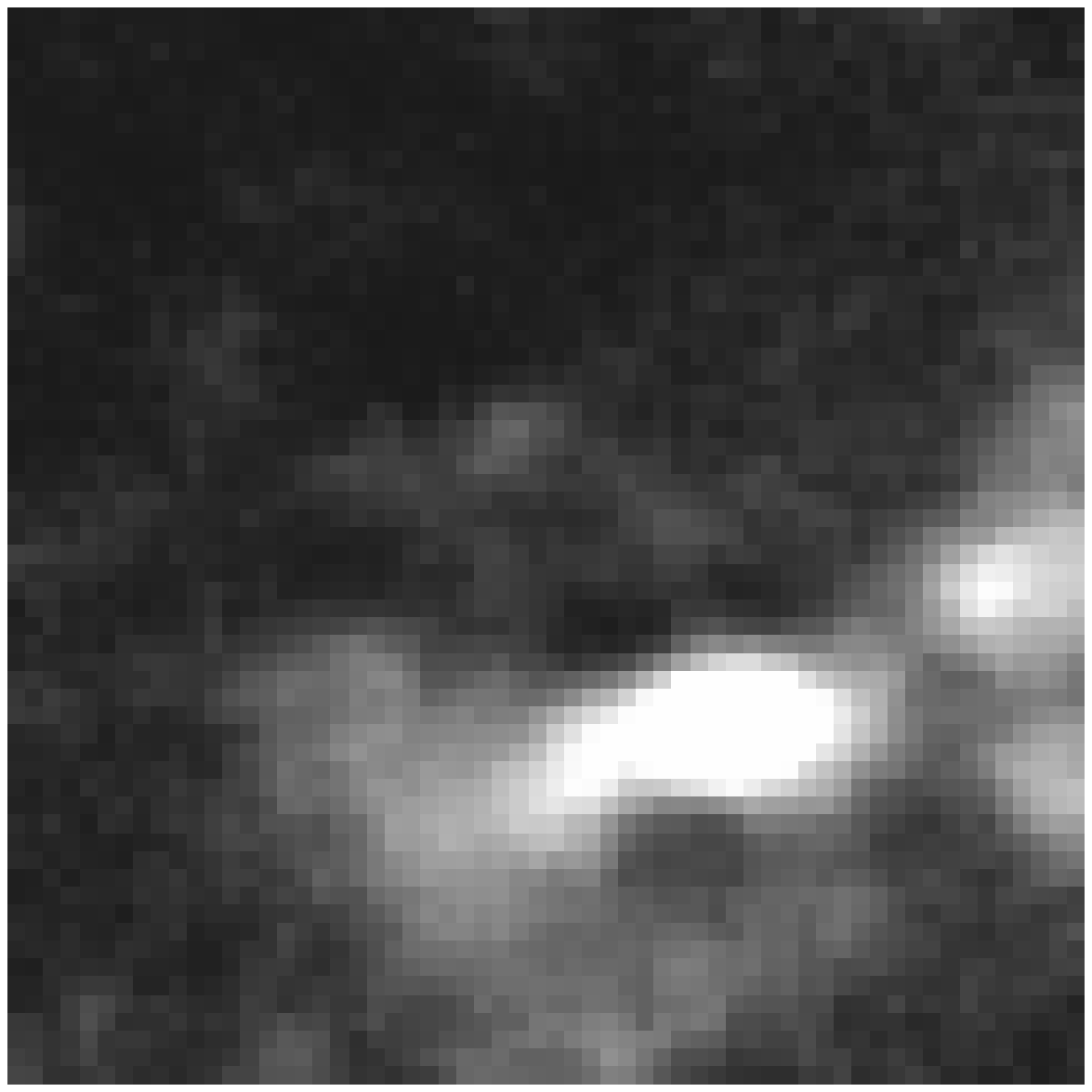}
    \hspace*{1.0cm}
    \FigureFile(40.0mm,40.0mm){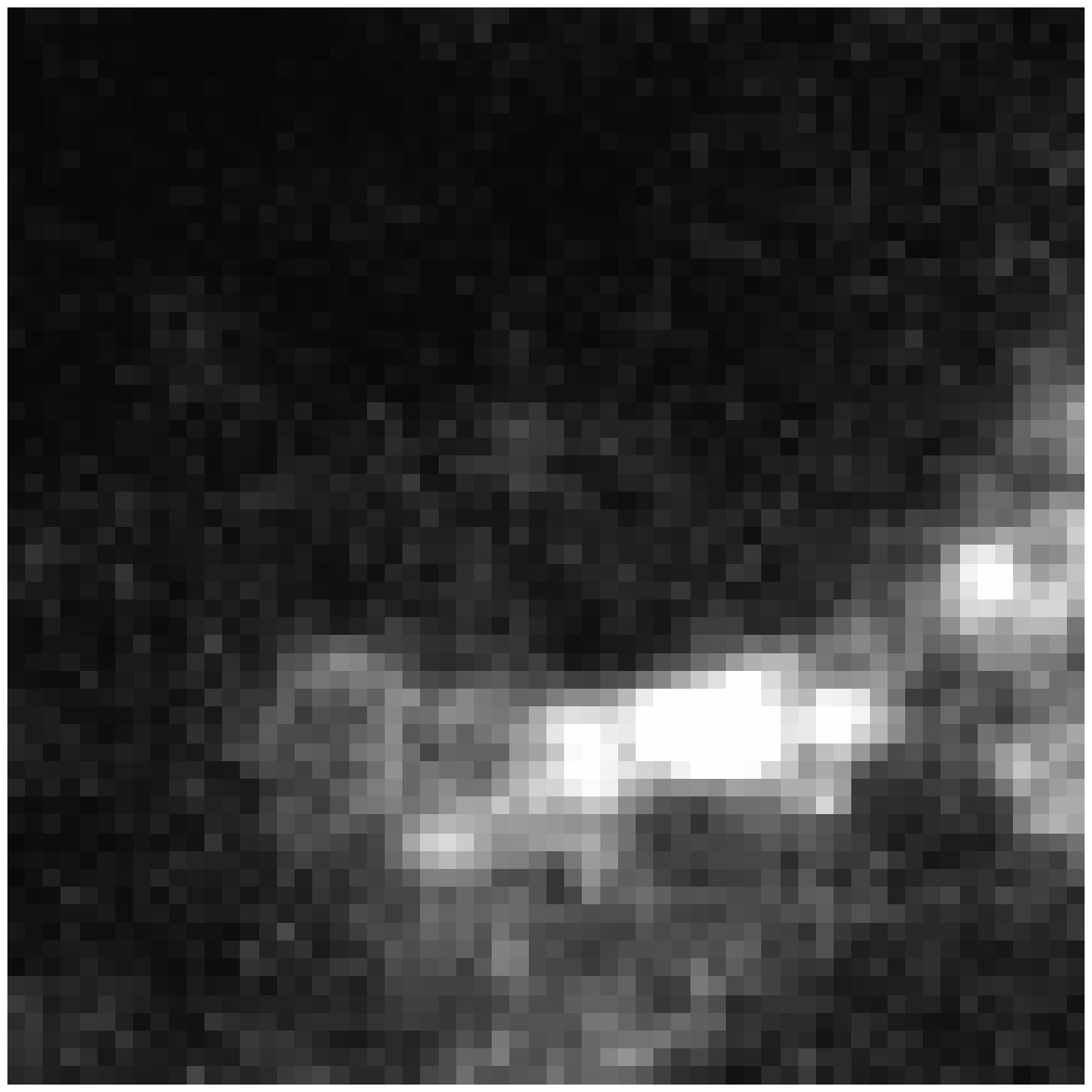}\\
    \vspace*{0.3cm}
    (d) Input \\
    \vspace*{-0.8cm}
    \FigureFile(120.0mm,96.0mm){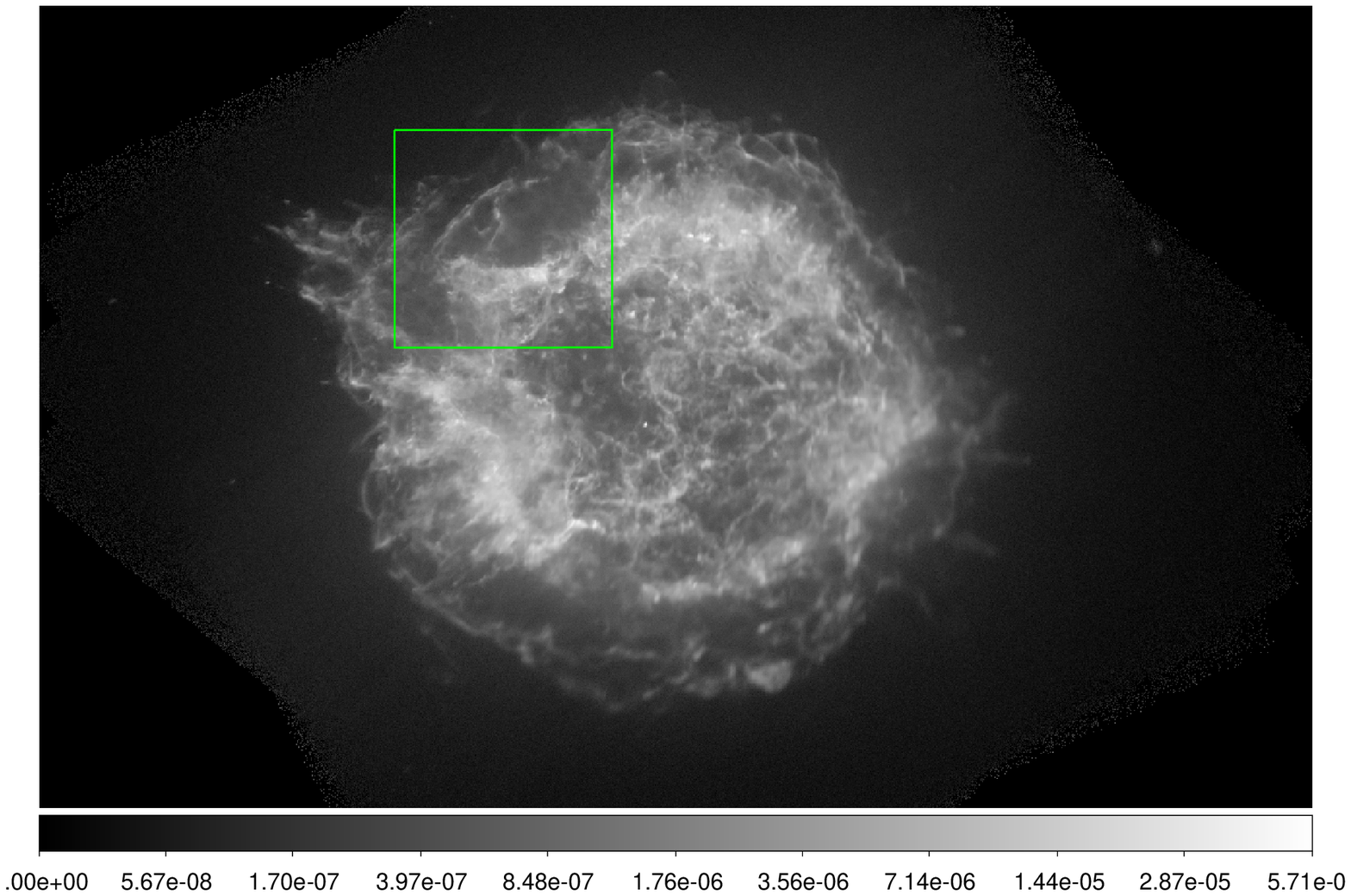}
  \end{center}
  \caption{Demonstration of image reconstruction of Cas A.  (a)
    Simulated image at focus (10M photons), (b) Same but at
    out-of-focus, (c) reconstructed images for 1M, 10M, 100M photons ,
    and (d) input image. The north-east squared region of the input
    image was simulated. The simulated image at the focus is shown in
    panel (a) whereas that at the out-of-focus positions is in panel
    (b). The images in panel (c) are reconstructed by using the images
    at out-of-focus.  The total number of input photons is
    $\sim 10^6$.  }
  \label{casa}
\end{figure}

For the reconstruction of more complex images, a larger number of photons is needed. 
Figure~\ref{casa} shows an example for the northeast part of Cas A, using the Chandra archived data . 
Using the response matrix ($T M$), we reconstructed the out-of-focus image , and show three images in figure~\ref{casa}.
The total number of their input photons are $\sim 10^6$, $\sim 10^7$, and $\sim 10^8$, respectively. The background is ignored. 
The shell structure of Cas A 
is identified in them. More detailed structures
would be identified if more photons could be collected.

\begin{figure}[ht!]
\begin{center}
\includegraphics[width=0.45\textwidth]{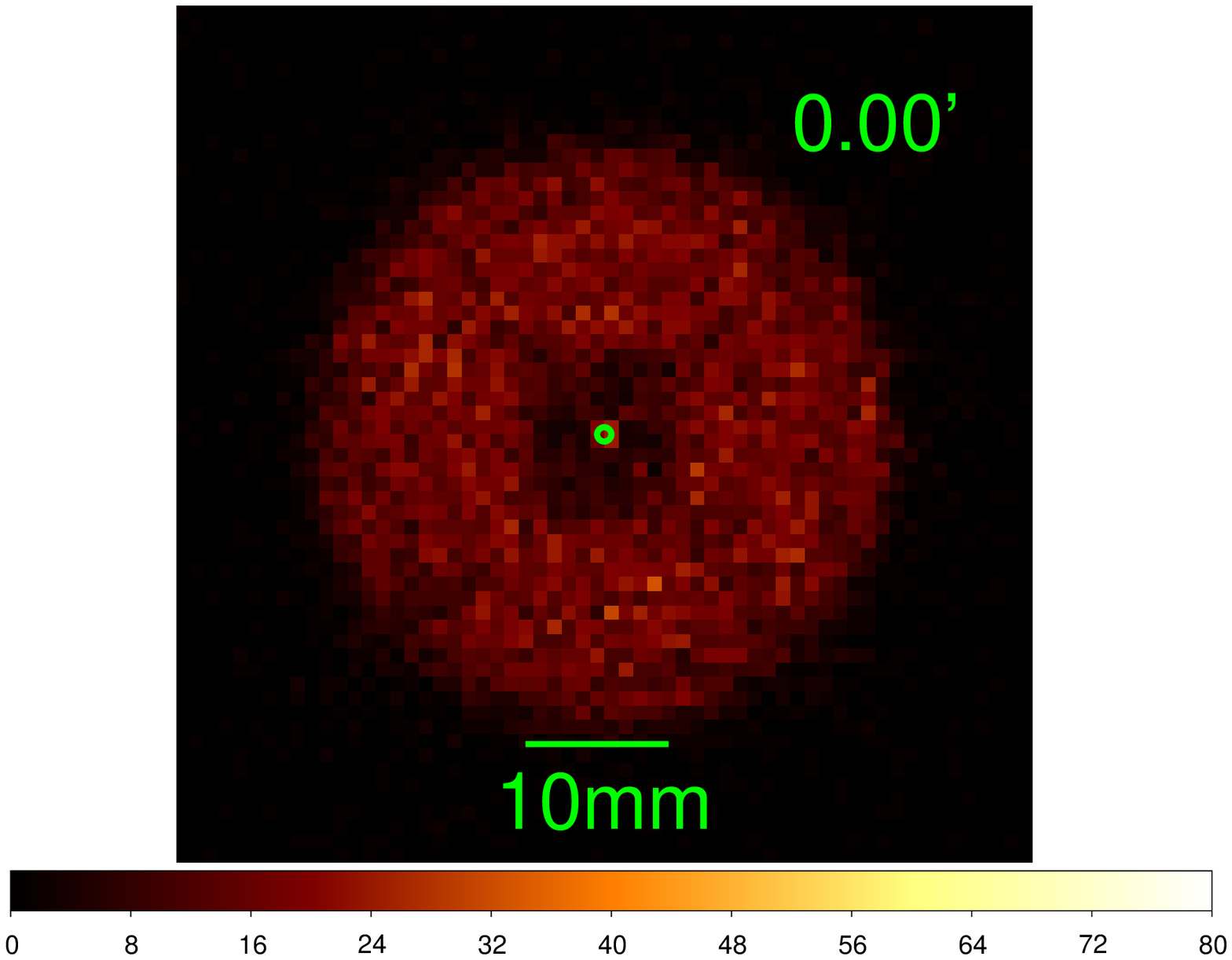}
\includegraphics[width=0.45\textwidth]{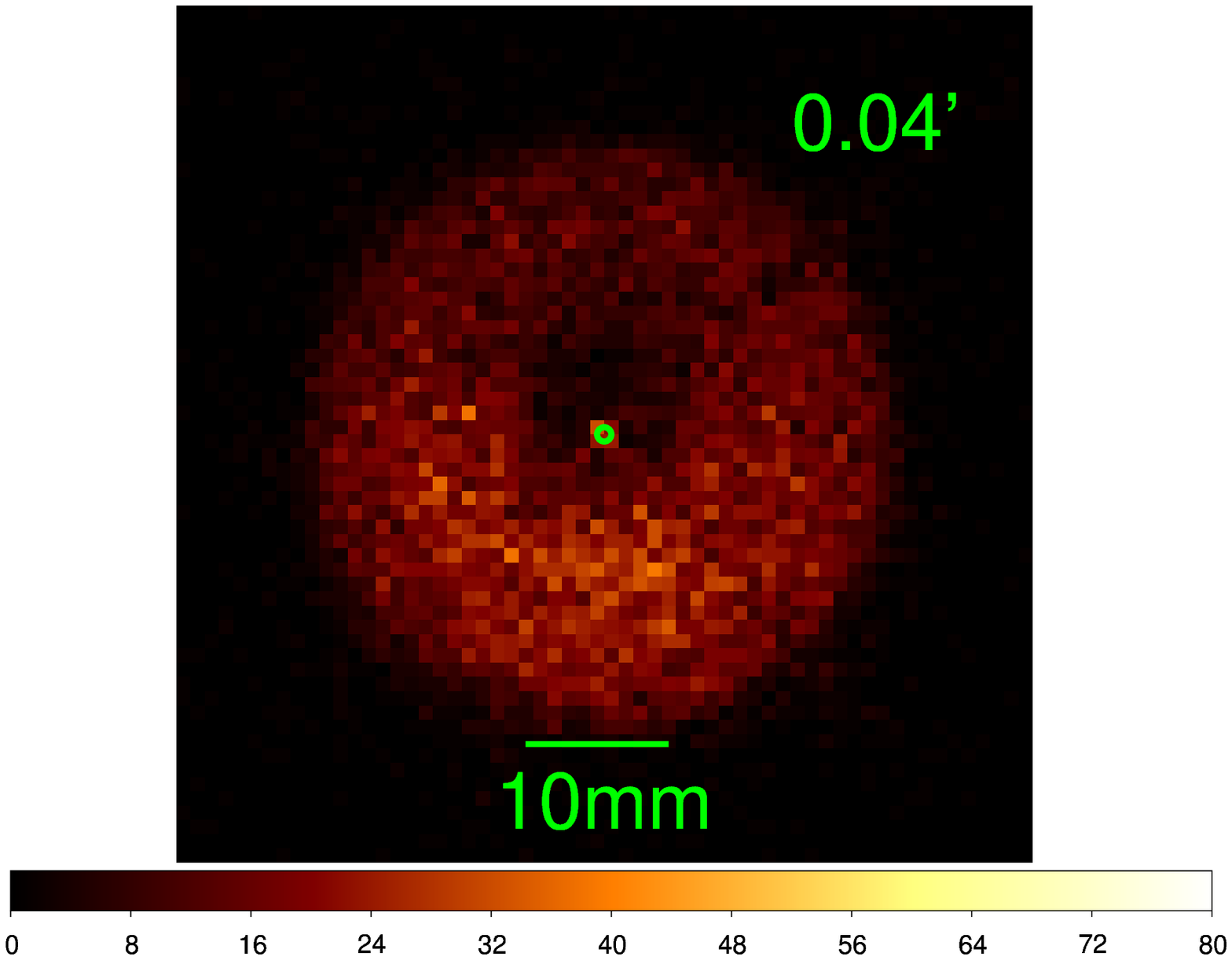}\\
\includegraphics[width=0.45\textwidth]{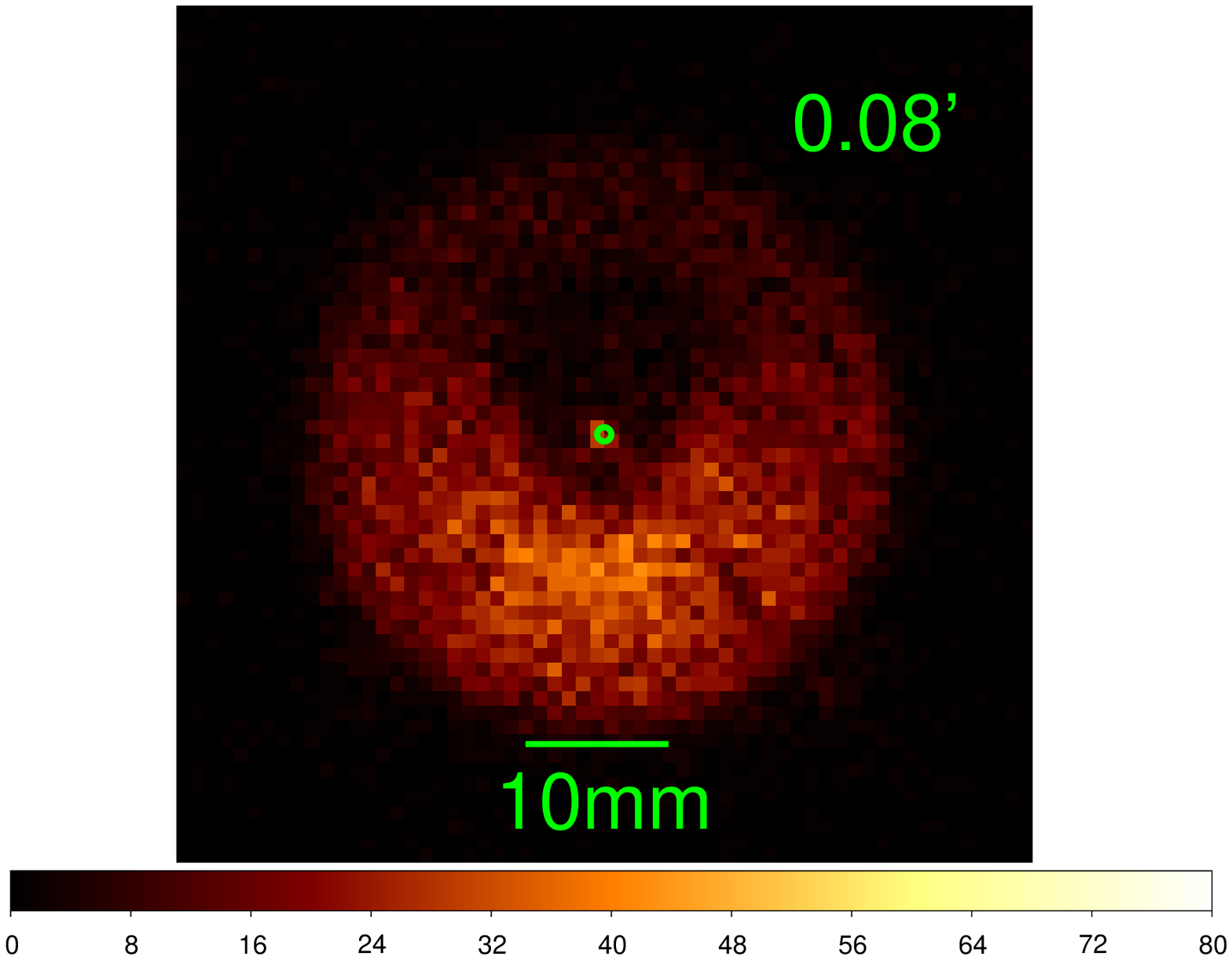}
\includegraphics[width=0.45\textwidth]{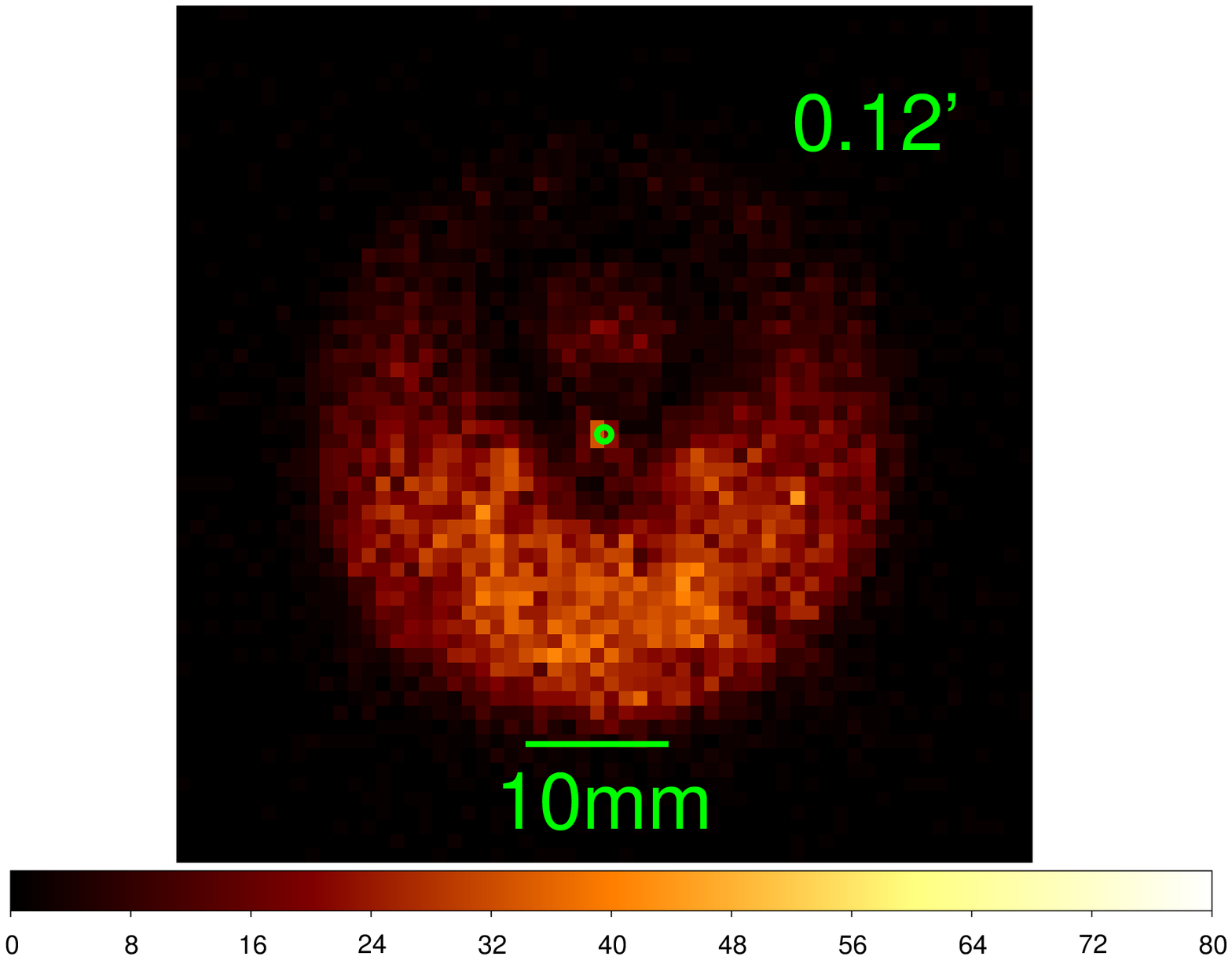}\\
\includegraphics[width=0.45\textwidth]{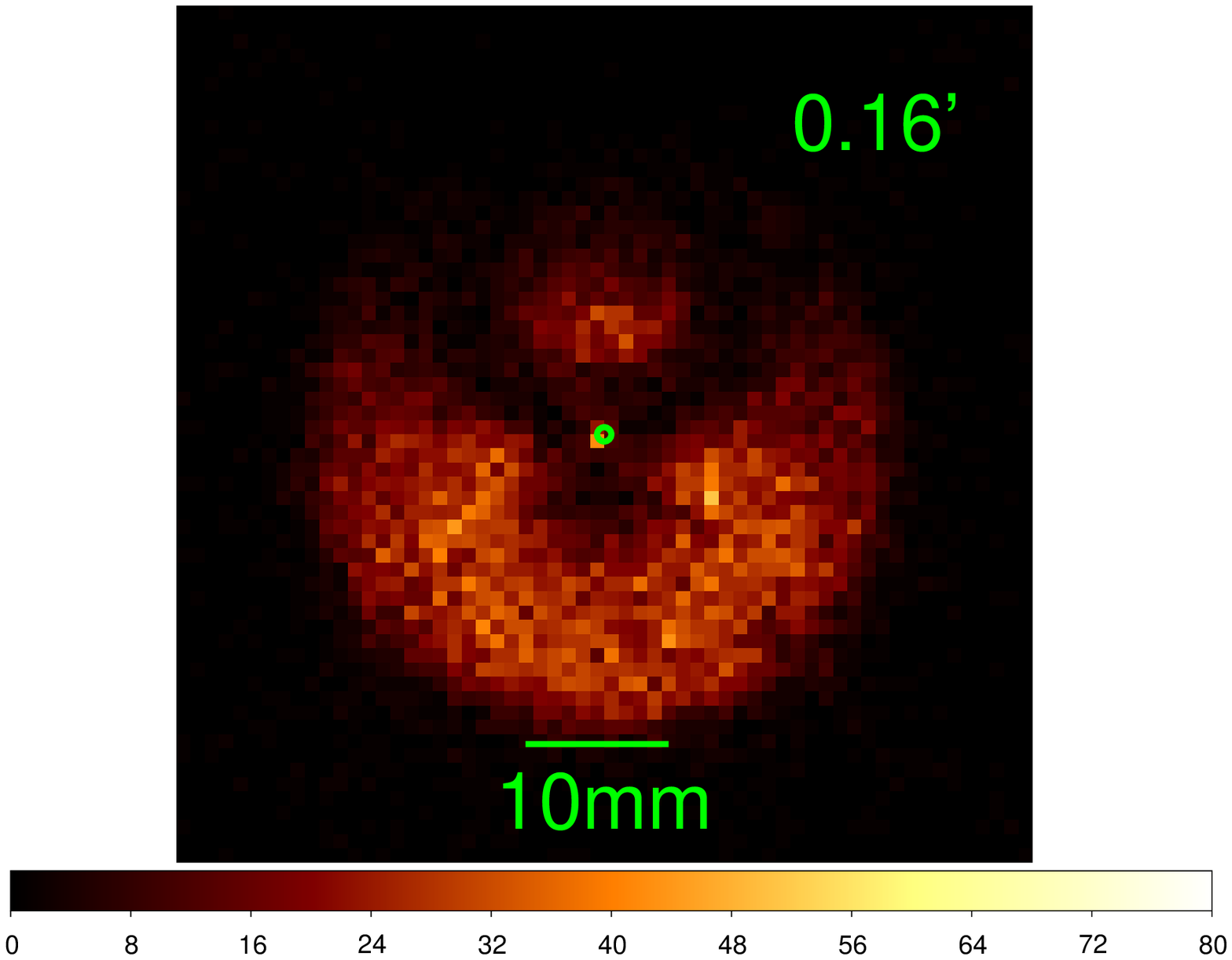}
\includegraphics[width=0.45\textwidth]{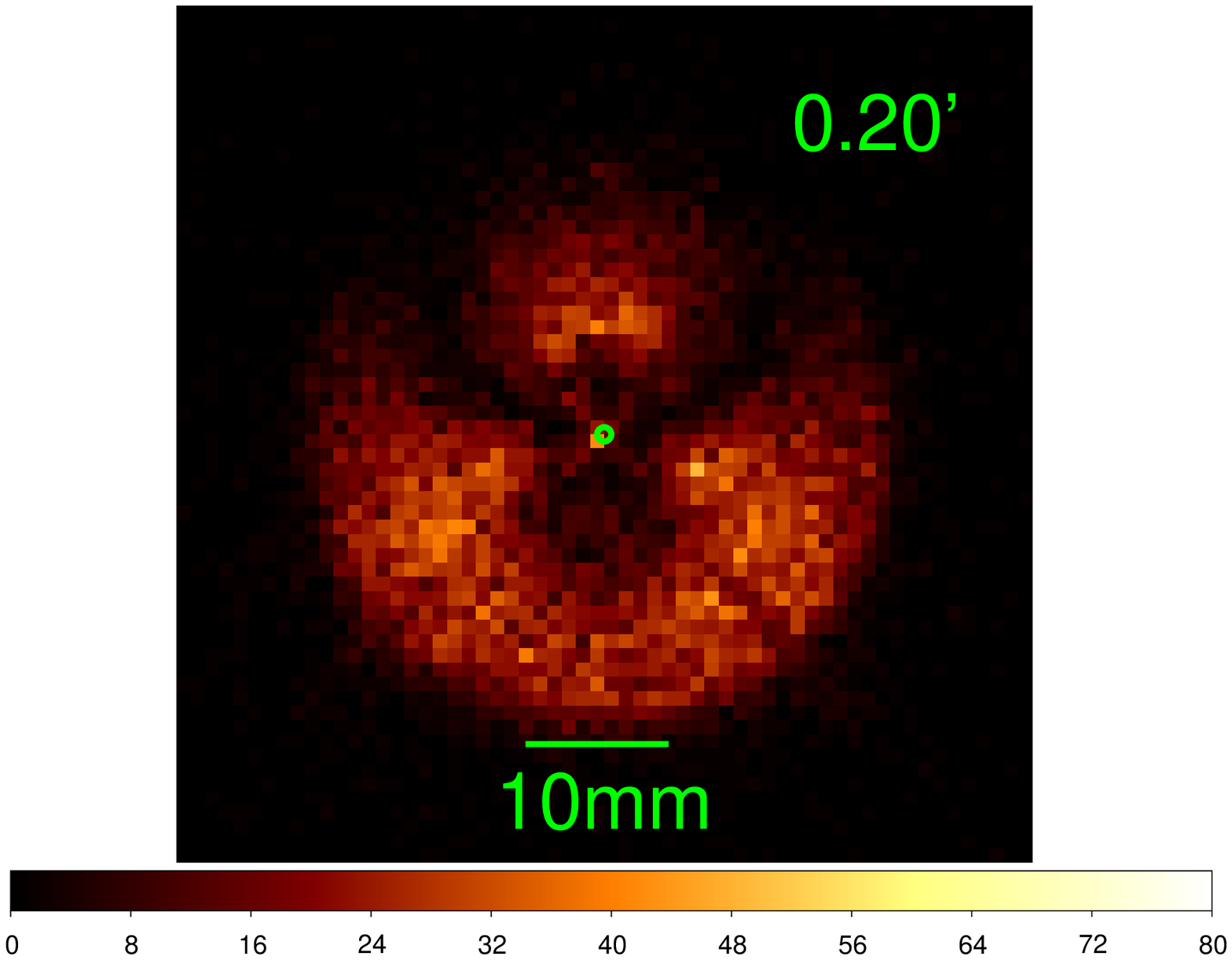}
 \end{center}
 \caption{Example of out-of-focus images of a rotation-mask pair\ (see figure~\ref{fig:slitexample}). 
The design parameters of the SXTs, which have an angular resolution of $\sim$1 arcmin, are adopted for this simulation. The detector is placed at $h=$500~mm from the focus.\ The off-axis angle of the input photons is labeled at the right-hand upper corner of each panel. 
}
\label{fig:example}
\end{figure}

\begin{figure}[ht!]
  \begin{center}
    \includegraphics[width=0.85\textwidth]
    {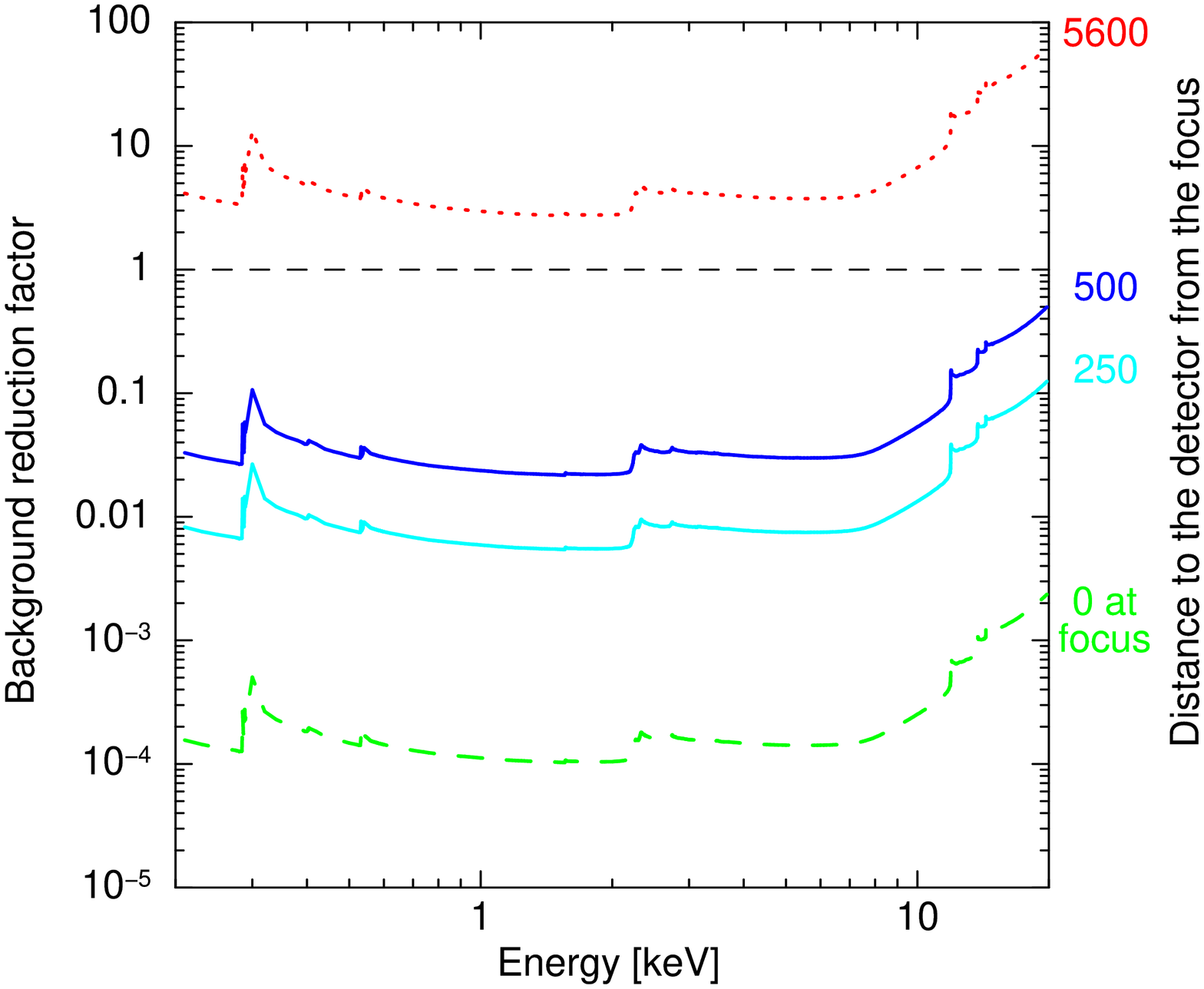}
  \end{center}
  \caption{ Reduction factor of the detector background, calculated
    with equation \ref{eqsen11} for out-of-focus positions at $h=250$,
    500, and 5,600~mm for cyan, blue, and red lines, respectively.
    The green dashed line shows the reciprocal of the throughput ratio
    of the focusing optics. The out-of-focus parameter $\alpha$ is
    $\sim$24 for $h=500$ mm and $\sim$12 for $250$ mm. The HPD of 1.2
    arcmins is assumed for $\Delta \theta_{\rm mirror}$.  }
  \label{fig:background}
\end{figure}

To demonstrate not only the angular resolution merit but also the sensitivity improvemnent, we calculate the comparison of the background-reduction factor, $f_{\rm bgd}$, to a point-like source along the on-axis (i.e., an offset angle of zero). 
This assumption is a good approximation for extended objects in arcmins such as Cas A since the field of view of the SXT-like mirror modules is $\sim15'$, and larger than an arcmin by an order. 
Figure~\ref{fig:background} shows the background-reduction factor, $f_{\rm bgd}$, due to the collimating effect of the mirror assembly at three out-of-focus positions.

 We use the following formula to calculate the reduction factor $f_{\rm bgd}$ at the focus: 
  \begin{eqnarray}                         
     f_{\rm bgd} =  (\frac{A_{\rm geometry}}{A_{\rm mirror}/2}) [\frac{\pi(\Delta \theta_{\rm mirror}/2)^2}{A_{\rm geometry}}]
     = \frac{\pi}{2} \frac{\Delta \theta_{\rm mirror}^2}{A_{\rm mirror}},
         \label{eqsen11}                            
 \end{eqnarray}      
 where the events are assumed to be accumulated over the area for the
 mirror angular resolution (HPD).

 In figure~\ref{fig:background}, the simulated telescope shows a
 signal-to-noise ratio better than the simple mask telescope by orders
 of magnitude. The factor will be improved more for a smaller
 out-of-focus distance $h$ or a smaller out-of-focus factor $\alpha$.
 Note that the effective area curve is simply scaled, the sensitivity
 of the focusing optics will be kept even if the mask unit is
 installed. It is known that certain numbers of photons are needed for
 reconstruction. The merit of our concept is in the sensitivity of the
 focusing optics. However, for imaging (i.e., reconstructing) , longer
 exposure by orders is necessary than for standard focusing optics.

\section{Conclusion}

We present a concept of the high angular resolution optics, using
focusing optics. In the proposed optics,
 
\begin{itemize}
\item A mask unit is installed in front of the focusing optics, which
  is the key to the achievements of a high angular resolution.  The
  angular resolution is improved when the two masks are placed at a
  larger distance between.  It is determined independently of the
  resolution of the focusing unit.

\item The pattern made with the masks is compressed by the focusing
  optics. The (two-dimensional) detector is positioned at the
  out-of-focus position. The out-of-focus image is then used to
  reconstruct the input image, using the inverse solution.  A moderate
  signal-to-noise ratio is obtained since the image is compressed.

\item The effective angular resolution of our proposed optics is
  dependent on the mask-pattern recognition uncertainty.  A less
  sufficiently recognized pattern can be originated from lower photon
  statistics, together with complex images of celestial objects and
  blurring of the out-of-focus images due to the angular resolution of
  the focusing optics.  Sufficient count-statistics are needed for the
  reconstruction.

\item A simple image, such as a double star, can be reconstructed with
  fewer photons. Complex images need much more photons to be
  reconstructed.

\item Any type of the two-dimensional detectors or detector array is
  in principle acceptable.  High angular-resolution spectroscopy can
  be performed, if the detector has a high energy resolution.  Finely
  spatially-resolved polarimetry is also quite possible with a
  suitable focal-plane detector.

\item Any type of focusing optics is in principle acceptable.
  Notably, a large effective area in a broad band can be achieved with
  reflective mirrors.  The bandpass can be extended to $\gamma$-ray if
  the Laue optics is used.
 \end{itemize}
 
\begin{ack}
  We thank the referees for their help in improving the quality of
  this paper.  The first and anonymous referee pointed out a critical
  error in an original manuscript.  We in part used the {\em ftools}
  software for the Suzaku and Hitomi missions released at {\rm
    https://heasarc.nasa.gov/lheasoft/ftools/}. The {\em ftools} is a
  general-purpose software package to manipulate FITS Files.  TS is
  supported by the Grant-in-Aid for Japan Society for the Promotion of
  Science (JSPS) Fellows (Grant Numbers JP16J03448 and JP19K14749). YM
  gratefully acknowledges funding from the Tanaka Kikinzoku Memorial
  Foundation. We thank M. Sakano (Wise Babel Ltd.) for English
  correction. We thank Y. Soong, T. Okajima, R. Asai and T. Dotani for
  discussions on many aspects of this optical system, including future
  applications.
\end{ack}
 
\bibliography{reference}

\end{document}